\documentclass[prl%
,twocolumngrid%
,superscriptaddress%
,floats
,aps%
]{revtex4}
\usepackage{amsmath}%

\begin{document}
\title{Jamming Model for the Extremal Optimization Heuristic} 
\date{\today}
\author{Stefan Boettcher}  
\email{sboettc@emory.edu}
\affiliation{Physics Department, Emory University, Atlanta, Georgia
30322, USA}  
\author{Michelangelo Grigni} 
\email{mic@mathcs.emory.edu}
\affiliation{Dept. of Mathematics and Computer Sciences, Emory
University, Atlanta, Georgia 30322, USA}

\begin{abstract} 
Extremal Optimization, a recently introduced meta-heuristic for hard
optimization problems, is analyzed on a simple model of jamming. The
model is motivated first by the problem of finding lowest energy
configurations for a disordered spin system on a fixed-valence
graph. The numerical results for the spin system exhibits the same
phenomenology found in all earlier studies of extremal optimization,
and our analytical results for the model reproduce many of these
features.  
\hfil\break  PACS number(s): 02.60.Pn, 05.40.-a, 64.60.Cn, 75.10.Nr.
\end{abstract} 
\maketitle

\section{Introduction}
\label{introduction}
Many situations in physics and beyond require the solution of NP-hard
optimization problems, for which the typical time needed to ascertain
the exact solution apparently grows faster than any power of the
system size~\cite{G+J}. Examples in the sciences are the determination
of ground states for disordered magnets~\cite{Pal,Hartmann,P+Y,EO_PRL}
or of optimal arrangements of atoms in a compound~\cite{Gould,B+S} or a
polymer~\cite{Erzan}. With the advent of ever faster computers, the
exact study of such problems has become
feasible~\cite{P+A,Rinaldi}. Yet, with typically exponential
complexity of these problems, many questions regarding those systems
still are only accessible via approximate, heuristic
methods~\cite{Rayward}. Heuristics trade off the certainty of an exact
result against finding optimal or near-optimal solutions with high
probability in polynomial time. Many of these heuristics have been
inspired by physical optimization processes, for instance, simulated
annealing~\cite{Science} or genetic algorithms~\cite{Holland}.

Extremal optimization (EO) was proposed recently~\cite{BoPe1}, and has
been used to treat a variety of combinatorial~\cite{GECCO,BGIP} and
physical optimization problems~\cite{EO_PRL}. Comparative studies with
simulated annealing~\cite{BoPe1,GECCO,EOperc} and other Metropolis
based heuristics~\cite{D+S} have established EO as a successful
alternative for the study of NP-hard problems, especially near phase
transitions~\cite{EOperc} that are associated with the most complex
instances of such
problems~\cite{Cheese,AI,Monasson,C+M,Zecchina,Franz}. Recently, EO
has also been successfully applied to Lennard-Jones
glasses~\cite{Gould}.

In this paper, we elucidate some properties of the EO algorithm with
analytical means. We motivate our theoretical model system with a
brief study of a disordered spin system on a random graph.
EO applied to finding ground states of
this system reveal the same generic properties found for the algorithm
previously. From this problem, we can abstract a set of evolution
equations which allow a complete analysis of EO as a function of its
single parameter, $\tau$, and the system size, $n$. In particular, an
optimal value for $\tau$ as a function of $n$ is determined in close
analogy with the scaling found numerically in all previous
studies~\cite{BoPe2}. We finish with a discussion of how this model
can be used also to investigate alternative versions of EO, or to
analytically compare EO with simulated annealing and other local
search heuristics.

\section{Spin glasses on fixed-valence random graphs}
\label{spin}
Disordered spin systems on random graphs have been investigated as
mean-field models of spin glasses~\cite{V+B} or optimization
problems~\cite{W+S,M+P,Zecchina,Franz}, since variables are long-range connected yet have
a small number of neighbors. Particularly simple are $\alpha$-valent
random graphs~\cite{M+P,Banavar,EOperc}. In these graphs each vertex
possesses a fixed number $\alpha$ of bonds to randomly selected other
vertices. Specifically, we have used the method described in
Ref.~\cite{Bollobas} to generate these graphs which are also referred
to as $\alpha$-regular graphs. (Note that self loops or double
connections are not allowed, and disconnected graphs are highly
unlikely). Just as on a lattice, one can assign a spin variable
$x_i\in\{-1,+1\}$ to each vertex, and couplings $J_{i,j}\in\{-1,+1\}$
to existing bonds between neighboring vertices $i$ and $j$. The energy
of the system then is the difference between violated bonds and
satisfied bonds,
\begin{eqnarray}
H=-\sum_{\{bonds\}} J_{i,j} x_i x_j.
\label{Heq}
\end{eqnarray}
It is more convenient to consider a linearly related quantity, which
merely tallies the number of violated bonds per spin in a configuration,
\begin{eqnarray}
e={H\over2n}+{\alpha\over4}\geq0,
\label{eeq}
\end{eqnarray}
where we have used the fact that each graph has a total of $\alpha
n/2$ bonds.

Clearly, for all $J_{i,j}\equiv1$ the spin system has two
ferromagnetic ground states with $e=0$ that are easy to find (all
$x_i=1$ or all $x_i=-1$). But for anti-ferromagnetic bonds
$J_{i,j}=-1$, the ground state energy depends on the disordered
structure of the graph itself. Only if all loops in the graph were of
even length (like in a hyper-cubic lattice), there are again simple
ground states, each with an alternating spin pattern (N\'eel
state). Instead, in a random graph, the disorder creates loops that
have an equal chance to be odd or even length. Thus, on average, half
of the loops can have all bonds satisfied, the other half will have at
least one bond frustrated.  Since the length of loops in random graphs
typically diverges with $\log(n)$, each odd loop almost certainly has
other odd loops as a neighbors to share a violated bond with.  In
fact, even for a spin glass, $J_{i,j}\in\{-1,+1\}$, the same argument
should hold, since only half of the loops will be frustrated and
neighboring frustrated loops can share violated bonds. We find that
the average ground state energies found for either bond distribution
are identical for $n\to\infty$, in support of the above argument, but
the results appear to differ in next-to-leading order corrections.

\begin{figure}[b!]
\vskip 4.0in  \includegraphics{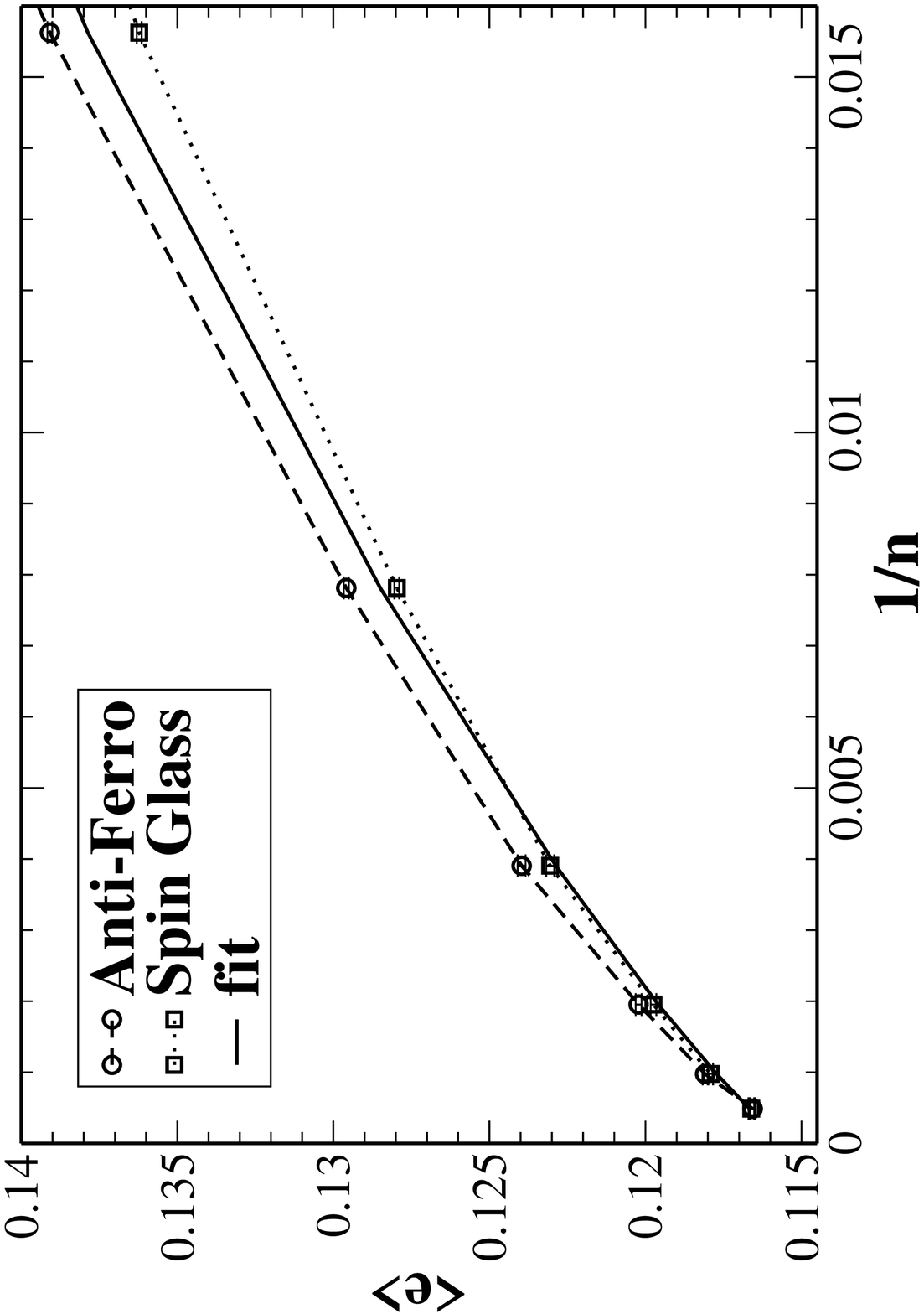} \includegraphics{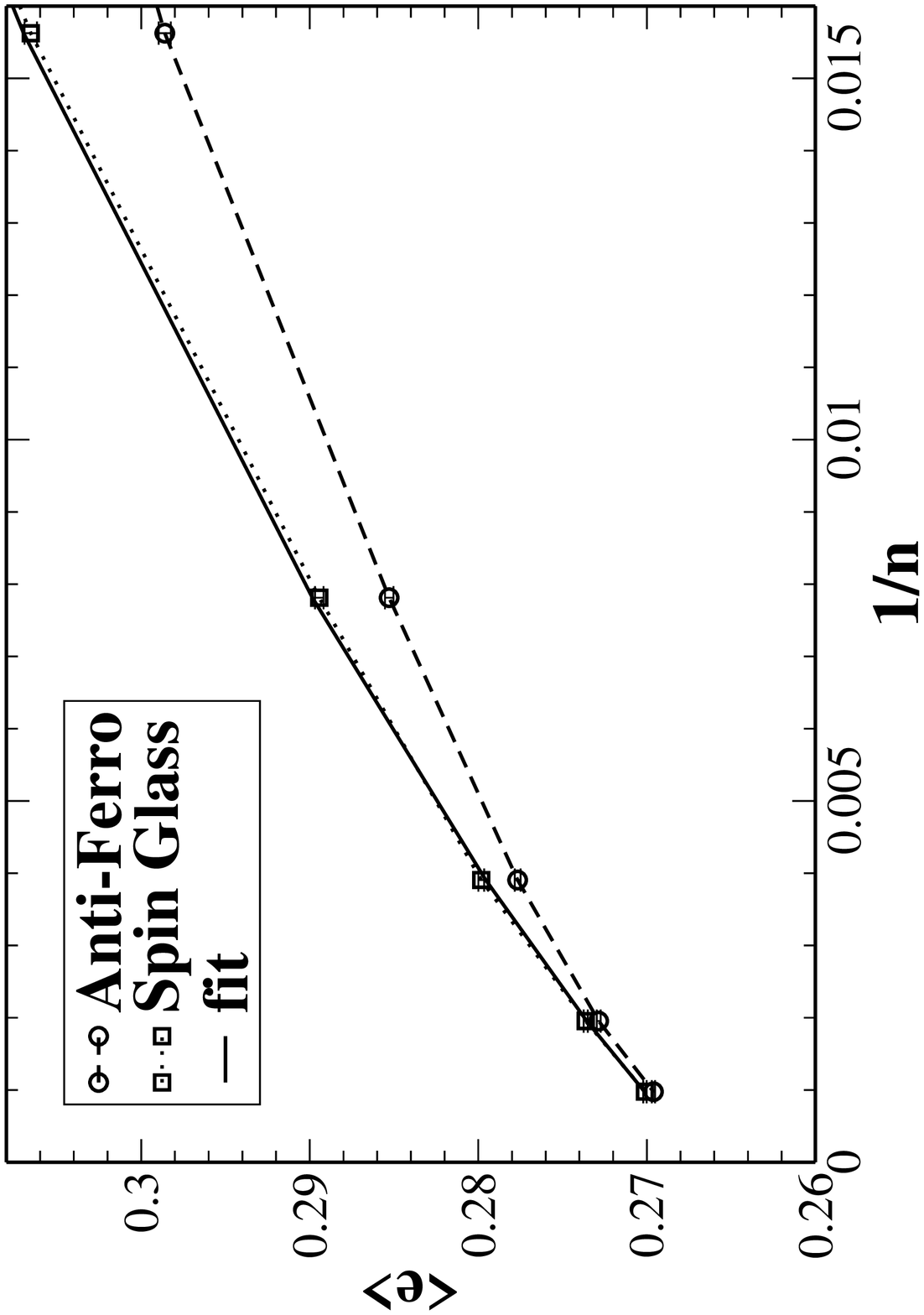}
\caption{Extrapolation for the number of violated bonds per spin, $e$,
as a function of $1/n$ (a) for trivalent and (b) for 4-valent graphs
of size $n=32,64,\ldots,1024$. Circles refer to an anti-ferromagnetic,
and the squares to a $\pm J$ bond distribution. The error bars for
$\langle e\rangle$  are smaller than the symbols. The data for the
spin glass is independent of the way the $\alpha$-valent graph was
formed and is best fit (continuous line) by
$e_{\alpha=3}(\infty)=0.1155(5)+0.35\ln(n)/n$ and
$e_{\alpha=4}(\infty)=0.266(1)+0.63\ln(n)/n$. We found that the data
for the anti-ferromagnet for smaller $n$ varies strongly with the way
the $\alpha$-valent graph was formed (here we used the method
described in Ref.~\cite{Bollobas}) and is difficult to fit. It is
apparent, though, that the difference between the data for the spin
glass and the anti-ferromagnet is decreasing for $n\to\infty$.  }
\label{extrapolation}
\end{figure}

\subsection{$\tau$-EO algorithm for $\alpha$-valent graphs}
To obtain the numerical results in Fig.~\ref{extrapolation}, we used
the following implementation of $\tau$-EO (see also
Ref.~\cite{EO_PRL}): For a given spin configuration on a graph, assign
to each spin $x_i$ a ``fitness''
\begin{eqnarray}
\lambda_i=-\#violated~bonds=-0,-1,-2,\ldots,-\alpha,
\end{eqnarray}
so that
\begin{eqnarray}
e=-{1\over2n}\sum_i\lambda_i
\label{lambdaeq}
\end{eqnarray}
is satisfied. Each spin falls into one of only $\alpha+1$ possible
states. Say, currently there are $n_{\alpha}$ spins with the worst
fitness, $\lambda=-\alpha$, $n_{\alpha-1}$ with $\lambda=-(\alpha-1)$,
and so on up to $n_0$ spins with the best fitness $\lambda=0$. (Note
that $n=\sum_in_i$.) Now draw a ``rank'' $k$ according to the
distribution
\begin{eqnarray} 
P(k)={\tau-1\over1-n^{1-\tau}} k^{-\tau}\quad(1\leq k\leq n).
\label{taueq}
\end{eqnarray}
Determine $0\leq j\leq\alpha$ such that
$\sum_{i=j+1}^{\alpha}n_i<k\leq\sum_{i=j}^{\alpha}n_i$. Finally,
select any one of the $n_j$ spins in state $j$ and reverse its spin
{\em unconditionally.} As a result, it and its neighboring spins
change their fitness. After all the effected $\lambda$'s and $n$'s are
reevaluated, the next spin is chosen for an update.

\begin{figure}[b!]
\vskip 2.0in  \includegraphics{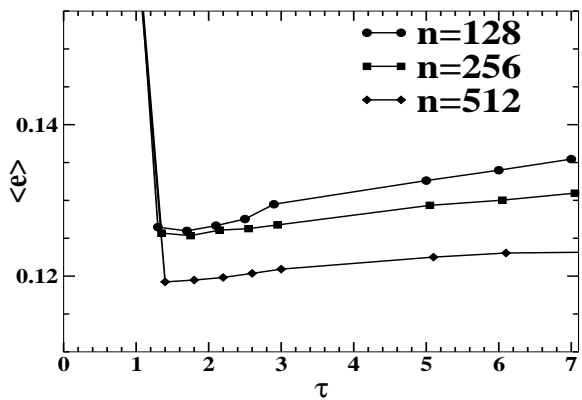}
\caption{Plot of the number of violated bonds per spin
$\left<e\right>$ as a function of $\tau$ as obtained by $\tau$-EO for
a $\pm J$-spin glass on trivalent graphs. Shown are the results for
$e$ averaged over 4 runs each on a set of 20 graphs for $n=128$, 256,
and 512. While the results clearly get worse rapidly for $\tau\leq1$,
even for $\tau\ll1$ a decline in the quality can be observed.
 (The
weak dependence of $e$ for large $\tau$ may indicate that a greedy
approach to finding ground states will yield good
approximations~\protect\cite{Moore+Palmer}.) 
Despite of the slow variation with
$\tau$, the value of $\tau_{\rm opt}$ where $\left<e\right>$ is
minimized clearly decreases toward $\tau=1^+$ with increasing $n$,
consistent with Eq.~(\protect\ref{tauscaleq}).  }
\label{tauplot}
\end{figure}

This EO implementation updates spins with a ($\tau$-dependent) bias
against poorly adapted spins on behalf of Eq.~(\ref{taueq}). This process is
``extremal'' in the sense that it focuses on atypical variables, and
it forms the basis of the EO method. The only adjustable parameter in
this algorithm is the power-law exponent $\tau$. For $\tau=0$,
randomly selected spins get forced to update, resulting in merely a
random walk through the configuration space. The search is ergodic but
yields poor results. For $\tau\to\infty$, only spins in the worst
state get updated which quickly traps the update process to a small
region of the configuration space which may be far from a near-optimal
solution. Ergodicity is broken in the sense that configurations far
from the initial conditions are unlikely to be reached within a given
runtime. The dependence of performance on $\tau$ for this and all
previous implementations of $\tau$-EO (for quite different
optimization problems~\cite{BoPe2,EO_PRL}) exhibits the features shown
in Fig.~\ref{tauplot}: The best average performance in approximating
ground state energies at a fixed runtime is obtained for a value of
$\tau_{\rm opt}$ slightly larger than $1$, and $\tau_{\rm opt}\to1^+$
for $n\to\infty$. (Note that $\tau=1$ would be a poor choice! Practical values for, say, an $n=1000$ random graph typically range from  $\tau_{\rm opt}\approx1.1$ for spin glasses to $\tau_{\rm opt}\approx1.6$ for bipartitioning.) In fact, the (more extensive) numerical data
presented in Ref.~\cite{BoPe2} suggested a simple argument that yields
\begin{eqnarray}
\tau_{\rm opt}\sim1+ {\ln\left({a\over\ln n}\right)\over\ln
n},\quad(n\to\infty,\ln(n)\ll a\ll n),
\label{tauscaleq}
\end{eqnarray}
where $t_{\rm max}=a\,n$ was used as the maximum number of
updates for a single EO run.  This asymptotic behavior was justified
by placing $\tau_{\rm opt}$ at the ``edge to ergodicity,'' a point
between having $\tau$ large enough to descent into local minima while
having $\tau$ just small enough to not get trapped inside the basin of
any local minimum. In the following we present a model to make this
notion more concrete.

\section{Evolution models}
\label{evolution}
We can abstract the random glass problem in Sec.~\ref{spin} into a
simple model which demonstrates previous observations about $\tau$-EO
in an analytically tractable way. Consider the spin system on an
$\alpha$-valent graph. Each spin $i$ can be in one of $\alpha+1$
states $\lambda_i$: either no adjacent bond is violated and $i$ is
among the $n_0$ spins, only one bond is violated placing it among the
$n_1$ spins, and so forth up to the $n_{\alpha}$ spins which have all
their adjacent bonds violated. Thus, one can define densities for each
of the $\alpha+1$ states, $\rho_i=n_i/n~(i=0,\ldots,\alpha)$. In
general, we can interpret any local search procedure, which only
updates a single variable at a time, simply as a set of evolution
equations for the $\rho_i(t)$, to wit
\begin{eqnarray}
{\dot\rho}_i=\sum_j T_{i,j}Q_j.
\label{rhodoteq}
\end{eqnarray}
Here, $Q_j$ is the probability that a spin in state $j$ gets updated,
and the matrix $T_{i,j}$ specifies the net transition to state $i$
given that a spin in state $j$ is updated.  Note that conservation of
probability requires
\begin{eqnarray}
\sum_j Q_j=1,
\label{Qnormeq}
\end{eqnarray}
and conservation of variables requires
\begin{eqnarray}
\sum_i T_{i,j}=0\quad (0\leq j\leq\alpha).
\label{Tnormeq}
\end{eqnarray}
Both, ${\bf T}$ and ${\bf Q}$ may generally depend on the $\rho_i(t)$
as well as on $t$ explicitly. (For instance, for simulated annealing
with a temperature schedule, the $Q_i$ could depend explicitly on $t$
through the changing temperature.)

Another relation is provided by the constraint
\begin{eqnarray}
\sum_i\rho_i(t)=1,
\label{rhonorm}
\end{eqnarray}
which implies $\sum_i{\dot\rho}_i=0$. Thus, one of the equations in
(\ref{rhodoteq}) is always redundant.  The cost per variable to be
minimized in Eq.~(\ref{eeq}) now reads
\begin{eqnarray}
e={1\over2}\sum_i i\rho_i\geq0,
\label{costeq}
\end{eqnarray}
with $e=0$ being optimal.

The advantage of this notation lies in the fact that the average
update {\em preference,} ${\bf Q}$, is separate from the update {\em
process} described by ${\bf T}$. For instance, for a random walk
(equivalent to $\tau$-EO at $\tau=0$ or simulated annealing at high
temperature) $Q_j(t)\equiv\rho_j(t)$, since the probability that a
spin in state $i$ gets chosen for an update is equal to the number of
those spins, no matter how that update is processed by ${\bf T}$. What
is typically unknown for a hard problem is the general form of ${\bf
T}$. But to understand the properties of a heuristic expressed through
${\bf Q}$, it may be revealing to ``design'' interesting ${\bf T}$.

\subsection{Annealed approximation to the glass problem}
We can construct ${\bf T}$ for the glass problem in Sec.~\ref{spin} on
a trivalent graph in an annealed approximation. Since ${\bf T}$ in
this case is quite messy, and of no great consequence beyond this
Section, we focus on one of its components, say, $T_{1,2}$. This
component represents the net flux in or out of $\rho_1$, given that a
variable in state $2$ gets updated. The annealed approximation
consists of the unbiased assumption that each of the $\alpha=3$
neighboring vertices can be in state $i$ with probability $\rho_i$
independently. Of course, no neighboring vertex can be in state $0$,
if the bond to it is violated, or in state $\alpha$, if the bond to it
is good.

For $T_{1,2}$, the vertex chosen for an update has $2$ violated bonds
and $\alpha-2=1$ good bond. First, when that vertex flips, there is a
shift of one variable (fraction $1/n$) from $\rho_2$ to $\rho_1$. The
neighboring vertex on the other end of each of the violated bonds
could be in state $1$, $2$, or $3$ with probability $\rho_1$,
$\rho_2$, or $\rho_3$, respectively, and the vertex attached via the
good bond could be in state $0$, $1$, or $2$ with probability
$\rho_0$, $\rho_1$, or $\rho_2$, respectively. Considering all allowed
combinations, we can find the relative (unnormalized) influx into any
of the $\rho_i$ as a consequence of updating the vertex at the
center. The sum of the influxes should equal the fraction of moved
vertices, $\alpha/n$, and the relative influxes can be normalized
accordingly. Finally, one can identify for each of the combinations
where that fraction of moved vertices originated from, which leads to
negative out-flux to the $T_{i,2}$ [which is obviously required to
satisfy Eq.~(\ref{Tnormeq})]. The out-flux out of state $i$ must be
proportional to $\rho_i$. Thus, we obtain the following three terms
contributing to $T_{1,2}$:
\begin{eqnarray}
T_{1,2}&=&{1\over n}+{\frac
{{\rho_0}\,{\rho_1}+2\,{\rho_1}\,{\rho_2}+{\rho_0}\,{\rho_3}+2\,{{\rho_2}}^{2}+3\,{\rho_0}\,{\rho_2}}{n\left
(1-{\rho_0}\right )\left (1-{\rho_3}\right )}}\nonumber\\ &&-{\frac
{\left (3\,{\rho_1}+3\,{\rho_2}+{\rho_3}+2\,{\rho_0}\right
){\rho_1}}{n\left (1-{\rho_0}\right )\left (1-{\rho_3}\right )}},
\label{T12eq}
\end{eqnarray}
and the construction of the other elements of ${\bf T}$ in this
annealed approximation proceeds equivalently.

It is not too hard to obtain some steady state (${\dot{\bf\rho}}=0$)
results for Eqs.~(\ref{rhodoteq}) with this particular ${\bf T}$,
supplemented by Eq.~(\ref{rhonorm}). One example would be the random
walk limit, $Q_i=\rho_i$, equivalent to $\tau=0$. More revealing for
the analysis of EO is the $\tau\to\infty$ limit. In that case, on each
update only one among the worst spins gets flipped. From some random
initial conditions, EO would empty out state $3$ first
($Q_3=1,~Q_2=Q_1=Q_0=0$), than empty out $2$, and so on, until a steady
state is reached with the highest non-empty state being $\rho_j$ with
some $j>0$. In this steady state, we can try to determine the
$\rho_i(\infty)$ with the ansatz ${\bar Q}_i=c_i$, $\sum_ic_i=1$,
where the average is taken over time. The only consistent balance is
obtained with state $3$ totally empty, $\rho_3(\infty)=0$ and $c_3=0$,
and state $2$ almost empty except for a single spin reaching the state
sometimes, i. e. $\rho_2(\infty)\approx0$ and $c_2>0$.  Hence,
$c_0=1-c_1-c_2$ and $\rho_1=1-\rho_0$, which leads to a drastically
simplified equations:
\begin{eqnarray}
0&=&c_2(3+2\rho_0)+c_1(2+\rho_0)-(1+3\rho_0),\nonumber\\
\medskip
0&=&c_2(1-4\rho_0)-c_1(1+2\rho_0)-(3-6\rho_0),\nonumber\\
\medskip
0&=&-c_2(3-2\rho_0)+c_1\rho_0+3(1-\rho_0),
\end{eqnarray}
all other equations being redundant. The solution is simply
\begin{eqnarray}
\rho_0(\infty)=1,&\quad&\rho_1(\infty)=0,\nonumber\\
\medskip
c_1={1\over2},&\quad& c_2={1\over2},
\end{eqnarray}
consistent with numerical simulation for all initial conditions. Thus,
in the steady state, almost all variables are in the ground state
except for a single vertex that is being bounced between state $1$ and
$2$.

The result that EO converges to the ground state for $\tau=\infty$,
while reassuring, is not very helpful to understand either EO or the
original problem. The annealed approximation has eliminated everything
that made the problem interesting, and EO's convergence for
$\tau=\infty$ to a perfectly optimized ground state clearly does not
resemble our numerical results from Sec.~\ref{spin}.

\subsection{Models with very simple flows}
\label{simpleflow}
Our naive annealed approximation has eliminated most of the relevant
features of the original, hard problem. Not surprisingly, it also
fails to predict the existence of a finite value for $\tau_{\rm opt}$
(see Fig.~\ref{tauplot}); it is easy to convince oneself that
$\tau=\infty$ is in fact the best case scenario for $\tau$-EO for all
initial conditions and even at finite runtime. Yet, two basic features
of the evolution equations remain appealing: (1) The behavior of a
system with a large number of variables can be abstracted into a
relatively simple set of equations, describing their dynamics with a
small set of unknowns, and (2) the separation of update process, ${\bf
T}$, and update preference, ${\bf Q}$, lends itself to an analytical
comparison between different heuristics. This distinction is possible,
of course, only as long as these heuristics can utilize the same
single-variable, local search process in ${\bf T}$. The question is:
Can we construct interesting processes ${\bf T}$ in the sense that
they capture salient features observed for local search on real,
NP-hard problems? We will show that even the most basic versions of
${\bf T}$ provide some insights into the workings of various local
search heuristics.

For simplicity, we choose $\alpha$ as small as possible for the three
following model situations. Without restriction of generality, in
these cases $\alpha=2$ is sufficient, but more complicated phenomena
could be accommodated with more states.  First, we consider the most
trivial case where a variable when updated merely moves from state $i$
to state $i-1$ for $i>0$, or from state $0$ to state $\alpha$ (to make
every state accessible),
$T_{i,j}=[-\delta_{i,j}+\delta_{i,(\alpha+j~{\rm
mod}~\alpha+1)}]/n$. This process is conveniently depicted as a flow
chart in Fig.~\ref{flowplot}a. Clearly, any gradient descent method
will be able to reach the ground state $e=0$ for this process, since
there are no barriers. For instance, simulated annealing with zero
temperature will reach this state in $O(n)$ trials, and $\tau$-EO for
$\tau=\infty$ will reach $e=0$ in $<n$ steps, when averaged over
initial conditions. [Note that in the above notation, $c_i=1/4$ solves
the steady state equations where $c_0>0$ implies $\rho_0(\infty)=1$,
$\rho_{i>0}(\infty)=0$.] Again, $\tau_{\rm opt}=\infty$ is obvious. In
fact, this model can be solved readily for any $\tau$ with the methods
to be developed below in Sec.~\ref{flowjam}. For the random walk
limit, $\tau=0$, it is $c_i=\rho_i(\infty)=1/4$ since
$Q_i\equiv\rho_i$.

Next, we can reverse the directions of transitions in the previous
example to obtain a less trivial case, which now possesses energetic
barriers. Here $T_{i,j}=[-\delta_{i,j}+\delta_{(\alpha+i~{\rm
mod}~\alpha+1),j}]/n$, as depicted in
Fig.~\ref{flowplot}b. Remarkably, the previous analysis for $\tau$-EO
(at least, for $\tau=0$ or $\infty$) does not change. The $e=0$ steady
state is reached again in $<n$ steps for $\tau=\infty$, since EO does
not reject uphill moves which are required here to arrive at state $0$
through state $2$, and $\tau_{\rm opt}=\infty$ again. On the other
hand, it is quite clear that simulated annealing will not arrive at
$e=0$ with finite probability in polynomial time, even for a
sophisticated temperature schedule. Such energetic barriers are, of
course, an inherent feature of many NP-hard problems, which makes this
simple model quite revealing.

\subsection{Model with jammed flow}
\label{flowjam}
\begin{figure}[t!]
\vskip 2.6in  \includegraphics{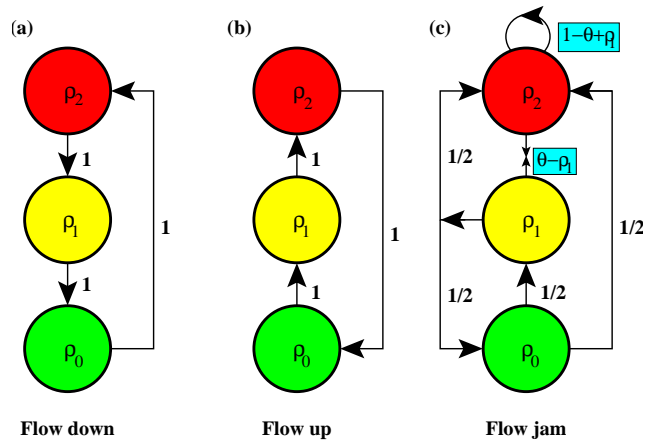}
\caption{ Plot of the flow diagrams for the different models discussed
in the text. Diagram (a) shows a situation in which variables in
higher states always evolve toward lower states (except for the lowest
state flowing up). In diagram (b), variables have to jump to higher
energetic states first before they can attain the lowest
state. Diagram (c) shows the model of a jam, where variables in the
highest state can only traverse through the intermediate state to the
lowest state, if the intermediate state moves its variables out of the
way first to keep its density $\rho_1$ below the threshold
$\theta$. The states have energies that increase from the bottom up,
the $\rho$'s mark the occupation density of each state, and arrows out
of a state indicate the rates $n\, T_{i,j}$ at which a variable flows
from state $j$ into another state $i$, {\em if\/} a variable in state
$j$ gets updated.  }
\label{flowplot}
\end{figure}

Naturally, the range of phenomena found in a local search of NP-hard
problems is not limited to energetic barriers. After all, so far we
have only considered constant entries for $T_{i,j}$. Therefore, in our
next model we want to consider the most simple case of ${\bf T}$
depending linearly on the $\rho_i$'s. Most of these cases reduce to
the phenomena already discussed in the previous examples. A entirely
new effect arises in the following case, also depicted in
Fig.~\ref{flowplot}c:
\begin{eqnarray}
 {\dot\rho}_0&=&{1\over  n}\left[-Q_0+{1\over2}Q_1\right],\nonumber\\
\medskip
 {\dot\rho}_1&=&{1\over n}\left[{1\over2}Q_0-Q_1+
(\theta-\rho_1)Q_2\right],\nonumber\\ \medskip  {\dot\rho}_2&=&{1\over
n}\left[{1\over2}Q_0+{1\over2}Q_1-
(\theta-\rho_1)Q_2\right],\nonumber\\ \medskip
1&=&\rho_0+\rho_1+\rho_2.
\label{thresheq}
\end{eqnarray}
Aside from the dependence of ${\bf T}$ on $\rho_1$, we have also
introduced the threshold parameter $\theta$. In fact, if
$\theta\geq1$, the model behaves effectively like the previous models,
and for $\theta\leq0$ there can be no flow from state $2$ to the lower
states at all. The interesting regime is the case $0<\theta<1$, where
further flow from state $2$ into state $1$ can be blocked for
increasing $\rho_1$, providing a negative feed-back to the system. In
effect, the model is capable of exhibiting a ``jam'' as observed in
many models of glassy dynamics~\cite{Jaeger,Ben-Naim,Ritort}, and
which is certainly an aspect of local search processes.

We proceed to calculate the unique fixed point of the system for EO with
arbitrary $\tau$. In the general case, the $Q$'s depend on the
$\rho$'s in a more complicated way. As described in the numerical
simulation of the glass on a random graph in Sec.~\ref{spin}, each
update a spin is selected based on its rank according to the
probability distribution in Eq.~(\ref{taueq}). When a rank $k(\leq n)$
has been chosen, a spin is randomly picked from state $\alpha$, if
$k/n\leq\rho_{\alpha}$, from state $\alpha-1$, if
$\rho_{\alpha}<k/n\leq\rho_{\alpha}+\rho_{\alpha-1}$, and so on. We
introduce a new, continuous variable $x=k/n$, approximate sums by
integrals, and rewrite $P(k)$ in Eq.~(\ref{taueq}) as
\begin{eqnarray} 
p(x)={\tau-1\over n^{\tau-1}-1} x^{-\tau}\quad\left({1\over n}\leq
x\leq 1\right),
\label{newtaueq}
\end{eqnarray}
where the maintenance of the low-$x$ cut-off at $1/n$ will turn out to
be crucial. Now, the average likelihood that a spin in a given state
is updated is given by
\begin{eqnarray}
Q_{\alpha}&=&\int_{1/n}^{\rho_{\alpha}} p(x)dx=
{1\over1-n^{\tau-1}}\left(\rho_{\alpha}^{1-\tau}-n^{\tau-1}\right),\nonumber\\
\medskip
Q_{\alpha-1}&=&\int_{\rho_{\alpha}}^{\rho_{\alpha}+\rho_{\alpha-1}}
p(x)dx\nonumber\\
&=&{1\over1-n^{\tau-1}}\left[\left(\rho_{\alpha-1}+
\rho_{\alpha}\right)^{1-\tau}-\rho_{\alpha}^{1-\tau}\right],\\
\medskip
&\ldots&\nonumber\\
\medskip
Q_0&=&\int_{1-\rho_0}^{1} p(x)dx={1\over1-n^{\tau-1}}
\left[1-\left(1-\rho_0\right)^{1-\tau}\right],\nonumber
\label{qeq}
\end{eqnarray}
where in the last line the norm $\sum_i\rho_i=1$ was used. These
values of the $Q$'s completely describe the update preferences for
$\tau$-EO at arbitrary $\tau$.

Inserting the set of Eqs.~(\ref{qeq}) for $\alpha=2$ into the model in
Eqs.~(\ref{thresheq}), we obtain
\begin{eqnarray}
 {\dot\rho}_0&=&{1\over
 n\left(1-n^{\tau-1}\right)}\left[-1+{3\over2}(1-\rho_0)^{1-\tau}-{1\over2}\rho_2^{1-\tau}\right],\nonumber\\
\medskip
 {\dot\rho}_1&=&{1\over
 n\left(1-n^{\tau-1}\right)}\Bigg[{1\over2}-{3\over2}(1-\rho_0)^{1-\tau}\nonumber\\
&&~~~~+\rho_2^{1-\tau}+(\theta-\rho_1)\left(\rho_2^{1-\tau}-n^{\tau-1}\right)\Bigg],\nonumber\\
 \medskip  {\dot\rho}_2&=&{1\over
 n\left(1-n^{\tau-1}\right)}\nonumber\\
&&~~~\left[{1\over2}-{1\over2}\rho_2^{1-\tau}-(\theta-\rho_1)\left(\rho_2^{1-\tau}-n^{\tau-1}\right)\right],\nonumber\\
 \medskip
1&=&\rho_0+\rho_1+\rho_2.
\label{floweq}
\end{eqnarray}
We abbreviate $A=(1-\rho_0)^{1-\tau}$ and $B=\rho_2^{1-\tau}$ to obtain for
the  steady state, ${\dot{\bf\rho}}=0$:
\begin{eqnarray}
0&=&-1+{3\over2}A-{1\over2}B,\nonumber\\
\medskip
0&=&{1\over2}-{3\over2}A+B+(\theta-\rho_1)\left(B-n^{\tau-1}\right),\nonumber\\
 \medskip
 0&=&{1\over2}-{1\over2}B-(\theta-\rho_1)\left(B-n^{\tau-1}\right),\nonumber\\
 \medskip  \rho_1&=&A^{1/(1-\tau)}-B^{1/(1-\tau)}.
\label{steadystateeq}
\end{eqnarray}
One of the first three equations is redundant, and we obtain
\begin{eqnarray}
0&=&{3\over2}(A-1)+\left[\theta-A^{1/(1-\tau)}+(3A-2)^{1/(1-\tau)}\right]\nonumber\\
&&~~~~~~~~~~~~~~~\left(3A-2-n^{\tau-1}\right),
\label{eigeneq}
\end{eqnarray}
where
\begin{eqnarray}
\rho_0&=&1-A^{1/(1-\tau)},\nonumber\\
\medskip
\rho_2&=&(3A-2)^{1/(1-\tau)},\nonumber\\
\medskip
\rho_1&=&1-\rho_0-\rho_2.
\label{alsoeigeneq}
\end{eqnarray}

\begin{figure}[b!]
\vskip 2.2in  \includegraphics{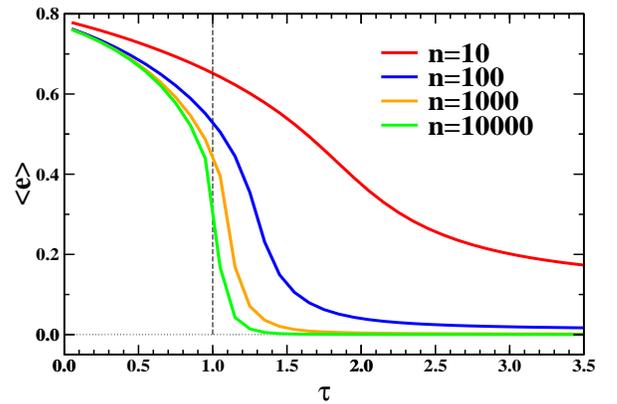}
\caption{Plot of $e=\sum_ii\rho_i/2$ as a function of $\tau$ resulting
from the solution of
Eqs.~(\protect\ref{eigeneq},\protect\ref{alsoeigeneq}) for
$\theta=1/2$ and various values of $n$. For $n\to\infty$, a sharp
transition emerges at $\tau=1$, giving optimal results $e\to0$ for all
$\tau>1$. But for $\tau>1$ this steady state is reached only for
suitable initial conditions, or after sufficient time, see
Fig.~\protect\ref{jamtauplot}.  }
\label{tauexactplot}
\end{figure}

The implicit Eq.~(\ref{eigeneq}) has some remarkable properties. It
has a single physical solution for the $\rho$'s for all
$0\leq\tau\leq\infty$, $0<\theta<1$~\cite{rem1}, and all $n$. In
particular, in the thermodynamic limit $n\to\infty$ a critical point
at $\tau=1$ emerges. If $\tau<1$, the $n$-dependent term in
Eq.~(\ref{eigeneq}) vanishes, allowing $A$, and hence the $\rho$'s, to
take on finite values, i. e. $e>0$. If $\tau>1$, the $n$-dependent
term diverges, forcing $A$ to diverge in kind, resulting in
$\rho_0\to1$ and $\rho_i\to0$ for $i>0$, i. e. $e\to0$. This behavior
of $e(\tau)$ for various $n$ is shown in Fig.~\ref{tauexactplot}.

Having a unique fixed point solution seems to be the last word on this
problem, with $\tau=\infty$ again being the most favorable value at
which the minimal energy $e=0$ is reached for sure. But it can be
shown that the system has an ever harder time to reach that point,
requiring typically $t=O(n^{\tau})$ update steps for a finite set of
initial conditions. Thus, for a given finite computational time
$t_{\rm max}$ the best results are obtained at some finite value of
$\tau_{\rm opt}$. In that, this model provides a new feature --- slow
variables impeding the dynamics of faster ones~\cite{PSAA} ---
resembling the observed behavior for EO on real problems, e.~g. the
effect shown in Fig.~\ref{tauplot}. In particular, this model provides
an analytically tractable picture for the relation between the value
of $\tau_{\rm opt}$ and the effective loss of ergodicity in the search
conjectured in Refs.~\cite{BoPe1,BoPe2}.

\begin{figure}[b!]
\vskip 2.2in  \includegraphics{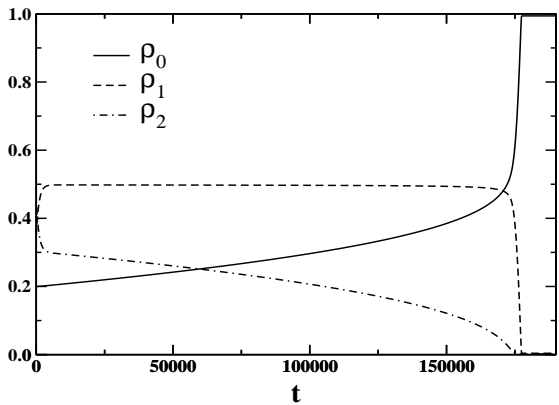}
\caption{Plot of the typical evolution of the system in
Eqs.~(\protect\ref{floweq}) for some generic initial condition that
leads to a jam. Shown are $\rho_0(t)$, $\rho_1(t)$, and
$\rho_2(t)$ for $n=1000$, $\tau=2$, $\theta=0.5$ and initial
conditions $\rho_0(0)=0.2$, $\rho_1(0)=0.35$, and
$\rho_2(0)=0.45$. Since $\rho_1(0)<\theta$, $\rho_1$ fills up to
$\theta$ almost instantly with variables from $\rho_2$ while $\rho_0$
stays constant. After that, $\rho_1\approx\theta$ for a very long time
($\gg n$) while variables slowly move down through state 1. Eventually,
after $t=O(n^{\tau})$, $\rho_2$ vanishes and EO can empty out $\rho_1$
directly which leads to the ground state $\rho_0=1$ ($e=0$) almost
instantly.  }
\label{evolutionplot}
\end{figure}

The generic evolution of the jamming model for $\tau>1$ is as follows:
If the initial conditions place a fraction $\rho_0>1-\theta$ already
into the lowest state, most likely no jam will emerge, since
$\rho_1(t)<\theta$ for all times, and the ground state is reached in
$<n$ steps. But if initially $\rho_1+\rho_2>\theta$, and $\tau$ is
sufficiently larger than unity, EO will drive the system to a
situation where $\rho_1\approx\theta$ by transferring variables from
$\rho_2$ to $\rho_1$. Then, further evolution becomes extremely slow,
delayed by the $\tau$-dependent, small probability that a variable in
state $1$ is updated ahead of all variables in state $2$. The typical
situation in that case is depicted in Fig.~\ref{evolutionplot}.

Clearly, the jam is {\em not} a steady state solution of the evolution
equations in (\ref{floweq}). [It is not even a meta-stable solution
since there are no energetic barriers. For instance, simulated
annealing at zero temperature would easily find the solution in
$t=O(n)$ without experiencing a jam. In reality, a hard problem would
most certainly contain combinations of jams, barriers, and possibly
other features.]  But Fig.~\ref{evolutionplot} suggest the right
asymptotic approach to evaluate the long-time behavior of the jam:
Consider that we start with initial conditions leading to a jam,
$\rho_1(0)+\rho_2(0)>\theta$. We can assume that
\begin{eqnarray}
\rho_1(t)=\theta-\epsilon(t)
\label{rho1eq}
\end{eqnarray}
with $\epsilon\ll1$ for $t\lesssim t_{\rm jam}$, where $t_{\rm jam}$
is the time at which $\rho_2$ gets small. To determine   $t_{\rm
jam}$, we apply Eq.~(\ref{rho1eq}) to the evolution equations in
(\ref{floweq}) to get
\begin{eqnarray}
 {\dot\rho}_0&\sim&{1\over
 n^{\tau}}\left[1-{3\over2}(1-\rho_0)^{1-\tau}+{1\over2}\rho_2^{1-\tau}\right],\nonumber\\
\medskip
0&=&{1\over2}-{3\over2}(1-\rho_0)^{1-\tau}+\rho_2^{1-\tau}-\epsilon
n^{\tau-1},\nonumber\\
\medskip
1&=&\rho_0+\theta+\rho_2.
\end{eqnarray}
Here, we have already dropped one of the equations (for
${\dot\rho}_2$) which was redundant. Now, we also can disregard the
equation containing $\epsilon$, its importance being that it
determines the first-order correction, $\epsilon=O(n^{1-\tau})$, {\em
consistently} as a function of the leading order contributions of
$\rho_0(t)$ and $\rho_2(t)$. Using the last (norm) equation and it's
derivative to leading-order, ${\dot\rho}_0=-{\dot\rho}_2$, we finally
obtain an equation solely for $\rho_2(t)$,
\begin{eqnarray}
 -{\dot\rho}_2&\sim&{1\over
  n^{\tau}}\left[1-{3\over2}(\theta+\rho_2)^{1-\tau}+{1\over2}\rho_2^{1-\tau}\right],
\end{eqnarray}
or, using the fact that $\rho_2$ almost instantly takes on the value
of $\rho_1(0)+\rho_2(0)-\theta=1-\theta-\rho_0(0)$ (see
Fig.~\ref{evolutionplot}),
\begin{eqnarray}
t\sim
n^{\tau}\int_{\rho_2(t)}^{1-\theta-\rho_0(0)}{2d\xi\over2-3(\theta+\xi)^{1-\tau}+\xi^{1-\tau}}.
\end{eqnarray}
We can estimate the duration of the jam, $t_{\rm jam}$, by setting
$\rho_2(t_{\rm jam})\approx0$, see Fig.~\ref{evolutionplot}:
\begin{eqnarray}
t_{\rm jam}\sim n^{\tau}f_{\tau}(1-\theta-\rho_0(0)),
\label{tjameq}
\end{eqnarray}
where we defined
\begin{eqnarray}
f_{\tau}(x)=\int_{0}^x{2d\xi\over2-3(\theta+\xi)^{1-\tau}+\xi^{1-\tau}}\quad(x\geq0).
\end{eqnarray}
Thus, the duration of the jam scales with $n^{\tau}$ times a constant
that depends on $\tau$, $\theta$, and the initial conditions.  As
stated before, if the initial conditions keep
$\rho_1(0)+\rho_2(0)<\theta$, most likely there will be no jam,
reflected in the fact that $f_{\tau}(x)$ goes to zero for $x\to0$. The
asymptotic scaling in Eq.~(\ref{tjameq}) conforms well with our
numerical simulations: with $f_2(0.3)\approx0.16$ and $n=1000$ we
obtain $t_{\rm max}\approx1.6\,10^5$, in good agreement with
Fig.~\ref{evolutionplot}.

The long-lived jams that occur for $\tau>1$ will have a significant
effect on the outcome of a local search with EO, which proceeds merely
with a finite runtime $t_{\rm max}$. For instance, for $t_{\rm
max}=O(n)$ there are always some initial conditions for which the jam
can not be resolved before $t_{\rm max}$, resulting in $e>0$. Thus,
the $\tau$-EO implementation faces two conflicting priorities: On one
side, larger $\tau$ increase the quality of the steady-state result
for $e$, away from the random-walk-like behavior at $\tau<1$, see
Fig.~\ref{tauexactplot}. On the other side, $\tau>1$ increases the
chance to get locked into a jam and never to reach that steady state
in finite runtime, see Fig.~\ref{evolutionplot}. In between these
conflicting interests, we find a preferred value for $\tau_{\rm opt}$
that averts both, the jam and the random walk, such that $\langle e
\rangle$, averaged over initial conditions, is minimized.

Let us assume we fix the runtime to be $t_{\rm max}=a\,n$, where $a$ is a
constant with $a\ll n$, so that $n<t_{\rm max}\ll n^{\tau}$ for
$\tau>1$. If we had chosen $\tau<1$ for our implementation, there are
no jams but we are sure to obtain less than optimal results for
$\langle e \rangle$ as in Fig.~\ref{tauexactplot}, so we will assume
$\tau>1$. In this case, we have to distinguish between three possible
outcomes to a single run of the EO algorithm, depending on the initial
conditions: (1) If $\rho_1(0)+\rho_2(0)<\theta$, the run will most
certainly reach the optimal state, $e=0$, within $t_{\rm max}$
updates, (2) even if $\rho_1(0)+\rho_2(0)>\theta$ but $t_{\rm
max}\gtrsim t_{\rm jam}$ from Eq.~(\ref{tjameq}), $e=0$ may be
reached. Only if (3) $\rho_1(0)+\rho_2(0)>\theta$ {\em and} $t_{\rm
max}\ll t_{\rm jam}$ are satisfied, the search will get stuck in a
state of $e>0$, with a value that depends on the initial
conditions. Averaging over all initial conditions, we find
\begin{eqnarray}
\left< e\right>&\approx&{1\over {\cal
N}}\int_0^1d\rho_0\,d\rho_1\,d\rho_2~
\delta\left(1-\rho_0-\rho_1-\rho_2\right){1\over2}\left(\sum_{i=0}^2i\rho_i\right)~\nonumber\\
\medskip
&&~~
u\left(1-\theta-\rho_0\right)~
u\left(f_{\tau}(1-\theta-\rho_0)n^{\tau}-t_{\rm max}\right),
\label{eavereq}
\end{eqnarray}
where $u(x)$ is the Heaviside step-function and $\delta(x)$ is the
Dirac delta-function. The norm is given by
\begin{eqnarray}
{\cal N}=
\int_0^1~d\rho_0\,d\rho_1\,d\rho_2~\delta\left(1-\rho_0-\rho_1-\rho_2\right)={1\over2}.
\end{eqnarray}
Hence, we obtain
\begin{eqnarray}
\left< e\right>\approx{3\over2}\int_0^{{\rm
max}\left\{0,1-\theta-f_{\tau}^{-1}\left(t_{\rm
max}/n^{\tau}\right)\right\}}~ d\rho_0~\left(1-\rho_0\right)^2.
\label{emineq}
\end{eqnarray}
The average energy $\left< e\right>$ in Eq.~(\ref{emineq}) will start
to rise for increasing $\tau$ as soon as the upper integration limit
becomes non-zero, or when $t_{\rm max}\approx
f_{\tau}(1-\theta)n^{\tau}$. If $t_{\rm max}\ll
f_{\tau}(1-\theta)n^{\tau}$, i.~e. for $\tau\gg1$, Eq.~(\ref{emineq})
predicts for the average energy $\left< e\right>=(1-\theta^3)/2$.

Since $\left< e\right>$ will reach its minimum value right before the
onset of jams cause its rise, we can use this relation to estimate the
optimal value of $\tau$. In effect, this justifies the connection
between $\tau_{\rm opt}$ and the ``edge of ergodicity'' noted in
Ref.~\cite{BoPe1}. Since the dependence of $f_{\tau}$ on $\tau$ is
much weaker than the exponential $n^{\tau}$, we can write
\begin{eqnarray}
\tau_{\rm opt}&\sim& {\ln\left(t_{\rm
max}/f_{\tau}(1-\theta)\right)\over\ln n},\nonumber\\
\bigskip
&\sim&1+ {\ln\left(a/f_{\tau}(1-\theta)\right)\over\ln n},
\label{tauopteq}
\end{eqnarray}
where we have used our choice $t_{\rm max}=an$. In recognition of the
fact that $\tau\to1^+$ for $n\to\infty$, we can simplify the last
expression by expanding $f_{\tau}(x)$ in that limit to get
\begin{eqnarray}
f_{\tau}(x)\sim
{2\over\tau-1}\int_0^x{d\xi\over\ln\left[{(\theta+\xi)^3\over\xi}\right]}\quad
(\tau\to1).
\label{fexpansioneq}
\end{eqnarray}
Note that the pole at $\tau=1$ is a generic consequence of
Eq.~(\ref{Tnormeq}), independent of the choice of the particular
$T_{i,j}$ depicted in Fig.~\ref{flowplot}c. If we insert
Eq.~(\ref{fexpansioneq}) into Eq.~(\ref{tauopteq}), we {\em exactly}
reproduce the $n$ dependence given in Ref.~\cite{BoPe2}, see
Eq.~(\ref{tauscaleq}) above. Numerical, we get at $\theta=1/2$ for
$f_{\tau}(1/2)\approx(2\ln2-1)/(\tau-1)$, and using $a=100$ and
$n=10$, 100, 1\,000, and 10\,000, Eq.~(\ref{tauopteq}) predicts
$\tau_{\rm opt}\approx3.5$, 2.1, 1.6, and 1.4.

\begin{figure}
\vskip 2.2in \includegraphics{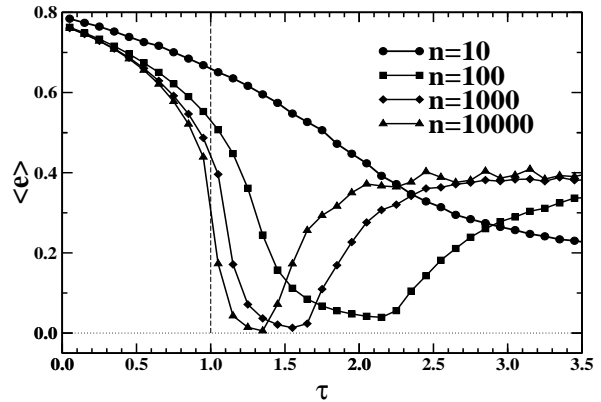}
\caption{Plot of the energy $\left< e\right>$ averaged over many
$\tau$-EO runs with different initial conditions as a function of
$\tau$ for $n=10$, 100, 1000, and 10000 and $\theta=1/2$. For small
values of $\tau$, $\left< e\right>$ closely follows the steady state
solutions plotted in Fig.~\protect\ref{tauexactplot}. It reaches a
minimum at a value near the prediction for $\tau_{\rm opt}\approx3.5$,
2.1, 1.6, and 1.4, and rises sharply beyond that. It reaches an
asymptotic value approaching the prediction of $\left< e\right>\approx0.44$ for $\tau\to\infty$.  }
\label{jamtauplot}
\end{figure}

We can compare this prediction with numerical simulations of $\tau$-EO
applied directly to the jamming system described in
Fig.~\ref{flowplot}c [not just the evolution equation
in~(\ref{floweq}) that use averaged probabilities $Q$]. In
Fig.~\ref{jamtauplot} we show the results for $\left< e\right>$ as a
function of $\tau$ for $n=10,~100,~1000,$ and 10000 at $t_{\rm
max}=100n$. Initially, for $\tau\lesssim1$, $\left< e\right>$ reaches
the steady state result from Fig.~\ref{tauexactplot} for any initial
condition. But, as predicted, $\left< e\right>$ reaches a minimum at a
$\tau_{\rm opt}$ beyond which it starts to rapidly deviate from the
steady state solution. This is the ``ergodic edge'' beyond which
unresolved jams effect the observed value of $\left< e\right>$. Our
prediction for $\tau_{\rm opt}$ appears to become increasingly
accurate for $n\to\infty$. Furthermore, for $\tau\to\infty$,
Eq.~(\ref{emineq}) predicts $\left< e\right>\sim7/16\approx0.44$ for
$\theta=1/2$, in reasonable agreement with the numerical value seen in
Fig.~\ref{jamtauplot}.

\section{Conclusion}
\label{conclusion}
We have presented a simple model to analyze the properties of local
search heuristics. This model was applied to extremal optimization and
we found conditions under which EO exhibited the same phenomenology on
the model as it does on real combinatorial optimization problems as
exemplified here by a frustrated spin system on a random graph. The
analytical results from the model in Eq.~(\ref{tauopteq}) closely
resemble Eq.~(\ref{tauscaleq}), the prediction from Refs.~\cite{BoPe1,BoPe2}.

Of course, the model is tailored more toward understanding the EO
mechanism and does not nearly represent all of the features of a hard
optimization problem. [After all, it takes EO only $O(n^{\tau})$
updates to find the ground state in the worst case for the model.]
Thus, finding a non-trivial value for $\tau_{\rm opt}$ in the model
merely provides an analogy. For instance, $\tau_{\rm opt}$ is somewhat
dependent on the relationship of $t_{\rm max}$ to $n$. If $t_{\rm
max}=O(n^l)$ then $\tau_{\rm opt}\sim l^+$ for $n\to\infty$ according
to the model. In this regard the analogy seems to hold in every
respect for the graph bipartitioning problem in
Refs.~\cite{BoPe1,BoPe2} where $t_{\rm max}\sim n$. But in our
numerical simulations for spin glass systems in Sec.~\ref{spin}
displayed in Fig.~\ref{tauplot}, or in Ref.~\cite{EO_PRL}, typically
$O(n^{3-4})$ updates were required to obtain consistent results for
increasing $n$, yet $\tau_{\rm opt}\to1^+$ was found irrespective.

We believe that our observation for the behavior of EO is quite robust
under variation of the entries for ${\bf T}$. More complicated choices
for ${\bf T}$ (which may be analytically less tractable) could be made
to represent hard problems more closely. In this sense, the separation
between ${\bf T}$ and ${\bf Q}$ allows to study comparisons between
different update modes, and even with other local search
procedures. As our examples in Figs.~\ref{flowplot}a-c show, simple
choices of ${\bf T}$ can lead to interesting scenarios, although there
is no real frustration. For instance, one could analyze the properties
of EO for different choices of the probability distribution over the
ranks in Eqs.~(\ref{taueq},\ref{newtaueq}).

It is more difficult to construct the $Q$'s for simulated
annealing. Let's assume that we consider a variable in state $j$ for
an update. Certainly, $Q_j$ would be proportional to $\rho_j$, since
variables are randomly selected for an update. The Boltzmann factor
$e^{-\beta\Delta E_j}$ for the potential update of a variable in $j$,
aside from the inverse temperature $\beta(t)$, only depends on the
entries for $T_{i,j}$:
\begin{eqnarray}
\Delta E_j&=&n\Delta e_j={n\over2}\left[\sum_ii\rho_i(t+1)-\sum_ii\rho_i(t)\right]_j,\nonumber\\
\medskip
&=&{n\over2}\sum_iiT_{i,j},
\end{eqnarray}
where the subscript $j$ expresses the assumption that a variable in
state $j$ is considered for an update. Hence, we find for the average
probability of an update of a variable in state $j$
\begin{eqnarray}
Q_j\propto\rho_j{\rm
min}\left\{1,\exp\left[-\beta\sum_iiT_{i,j}\right]\right\},
\end{eqnarray}
which is still short of a proper normalization. Similarly, comparisons
with other methods such as threshold annealing~\cite{FHS} can be
considered.

We would like to acknowledge helpful discussions with participants of
the 2001 Telluride workshop on energy landscapes, in particular Allon
Percus and Paolo Sibani. Special thanks to Remi Monasson whose visit
to Emory inspired the idea.


\begin{thebibliography}{}

\bibitem{G+J} M. R. Garey and D. S. Johnson,  {\it Computers and
Intractability, A Guide to the Theory of NP-Completeness}
(W. H. Freeman, New York, 1979).


\bibitem{Pal} K. F. Pal,
Physica A {\bf 223}, 283-292 (1996), and  {\bf 233}, 60-66 (1996).

\bibitem{Hartmann} A. K. Hartmann,
Phys. Rev. B {\bf 59}, 3617-3623 (1999), and  Phys. Rev. E {\bf 60},
5135-5138 (1999).

\bibitem{P+Y} M. Palassini and A.~P. Young,
Phys. Rev. Lett. {\bf 85}, 3017 (2000).

\bibitem{EO_PRL} S. Boettcher and A. G. Percus,
Phys. Rev. Lett. {\bf 86}, 5211-5214 (2001).

\bibitem{Gould} D.~L.~Blair and H.~Gould, private communication.

\bibitem{B+S} K.~K. Bhattacharya and J.~P. Sethna,
Phys. Rev. E {\bf 57}, 2553 (1998).

\bibitem{Erzan} E. Tuzel and A. Erzan,
Phys. Rev. E {\bf 61}, R1040 (2000).

\bibitem{P+A} R. G. Palmer and J. Adler,  Int. J. Mod. Phys. C {\bf
10}, 667 (1999).

\bibitem{Rinaldi} C. Desimone, M. Diehl, M. J\"unger, P. Mutzel,
G. Reinelt, G. Rinaldi,
J. Stat. Phys. {\bf 80}, 487-496 (1995).


\bibitem{Rayward} {\em Modern Heuristic Search Methods,}
Eds. V. J. Rayward-Smith, I. H.  Osman, and C. R. Reeves (Wiley, New
York, 1996).


\bibitem{Science} S. Kirkpatrick, C. D. Gelatt, and M. P. Vecchi,
Science {\bf 220}, 671-680 (1983).

\bibitem{Holland} J.~Holland, {\em Adaptation in Natural and
Artificial Systems\/} (University of Michigan Press, Ann Arbor, 1975).

\bibitem{BoPe1}  S. Boettcher and A. G. Percus,
Artificial Intelligence {\bf 119}, 275-286 (2000).

\bibitem{GECCO} S. Boettcher and A. G. Percus,
in {\it GECCO-99: Proceedings of the Genetic and Evolutionary
Computation Conference} (Morgan Kaufmann, San Francisco, 1999),
825-832.

\bibitem{BGIP} S. Boettcher, M. Grigni, G. Istrate, and A. G. Percus,
{\em Phase Transitions and Computational Complexity,} in preparation.


\bibitem{EOperc} S. Boettcher,
J. Math. Phys. A {\bf 32}, 5201-5211 (1999).

\bibitem{D+S} J. Dall and P. Sibani
Computer Physics Communication {\bf 141}, 260-267 (2001).


\bibitem{Cheese} P. Cheeseman, B. Kanefsky, and W.~M. Taylor,
in Proc. of IJCAI-91, eds. J. Mylopoulos and R. Rediter (Morgan
Kaufmann, San Mateo, CA, 1991), pp. 331--337.

\bibitem{AI} See {\em Frontiers in problem solving: Phase transitions
and complexity,} Special issue of Artificial Intelligence {\bf
81}:1--2 (1996).


\bibitem{Monasson} R. Monasson, R. Zecchina, S. Kirkpatrick,
B. Selman, and L. Troyansky,
Nature {\bf 400}, 133-137 (1999), and Random Struct. Alg. {\bf 15},
414-435 (1999).

\bibitem{C+M} S. Cocco and R. Monasson,
Phys. Rev. Lett. {\bf 86}, 1654-1657 (2001), and Europhys. Jour. B {\bf 22}, 505-531 (2001).


\bibitem{Zecchina}
F.~Ricci-Tersenghi, M.~Weigt, and R.~Zecchina,
Phys. Rev. E {\bf 63}, 026702 (2001).

\bibitem{Franz}
S.~Franz, M.~Leone, F.~Ricci-Tersenghi, and R.~Zecchina R,
Phys. Rev. Lett. {\bf 87}, 7209-7212 (2001).
 
\bibitem{BoPe2} S. Boettcher and A. G. Percus,
Phys. Rev. E {\bf 64} 026114 (2001).

\bibitem{V+B} L. Viana and A.~J. Bray,  J. Phys. C {\bf 18}, 3037
(1985).

\bibitem{W+S} K. Y. M.  Wong and D. Sherrington,
J. Phys. A: Math. Gen {\bf 20}, L793 (1987);

\bibitem{M+P} M. Mezard and G. Parisi,
Europhys. Lett. {\bf 3}, 1067 (1987).


\bibitem{Banavar} J. R. Banavar, D. Sherrington, and N. Sourlas,
J. Phys. A: Math. Gen {\bf 20}, L1 (1987).

\bibitem{Bollobas} B. Bollobas, {\it Random Graphs,} (Academic Press,
London, 1985).

\bibitem{Moore+Palmer} C. Moore and R. G. Palmer, 
{\em Upper and Lower Bounds on Spin-Glass Ground State Energies}
in preparation.

\bibitem{Jaeger} H.~M. Jaeger, S.~R. Nagel, R.~P. Behringer,
Rev.~Mod.~Phys {\bf 68} 1259-1273 (1996).

\bibitem{Ben-Naim} E. Ben-Naim, J.~B. Knight, E.~R. Nowak,
H.~M. Jaeger, and S.~R. Nagel,
Physica D {\bf 123}, 380-385 (1998).

\bibitem{Ritort} F. Ritort,
Phys.~Rev.~Lett. {\bf 75}, 1190-1193 (1995).

\bibitem{rem1} Actually, for $\theta\leq0.385$, there is a transition
to having three solutions near $\tau=1$, resulting in a first-order
transition for $n\to\infty$. We will only discuss $\theta>0.385$ here.

\bibitem{PSAA} R. G. Palmer, D. L. Stein, E. Abrahams, and
P. W. Anderson, Phys.~Rev.~Lett. {\bf 53}, 958 (1984).

\bibitem{FHS} A. Franz, K.-H. Hoffmann, P. Salamon,
Phys.~Rev.~Lett. {\bf 86} 5219-5222 (2001).
\end{thebibliography}
\end{document}